\documentclass[cameraready]{Interspeech}
\usepackage{booktabs}
\usepackage{multirow}
\usepackage{graphicx,algorithm,algorithmic}
\usepackage{pifont}  

\title{From Speech to Profile: A Protocol-Driven LLM Agent for Psychological Profile Generation}

\author[affiliation={1}, equalcontribution]{Xingjian}{Yang}
\author[affiliation={2,3}, equalcontribution]{Yudong}{Yang}
\author[affiliation={2,3}]{Zhixing}{Guo}
\author[affiliation={4}]{Yongjie}{Zhou}
\author[affiliation={2,3}]{Nan}{Yan}
\author[affiliation={2,3},correspondingauthor]{Lan}{Wang}

\address{
    $^1$ School of Science\&Engineering, South China University of Technology, China\\
    $^2$ Shenzhen Institutes of Advanced Technology, Chinese Academy of Sciences, China \\
    $^3$ Key Laboratory of Biomedical Imaging Science and System, Chinese Academy of Sciences, China \\ 
    $^3$ Shenzhen Mental Health Center\&Shenzhen Kangning Hospital, China 
}

\email{202330461911@mail.scut.edu.cn, yd.yang2@siat.ac.cn, lan.wang@siat.ac.cn}


\usepackage{comment}

\begin{document}
\maketitle

\begin{abstract}
The psychological profile that structurally documents the case of a depression patient is essential for psychotherapy. Large language models can be applied to summarize the profiles from counseling speech, however, it may suffer from long-context forgetting and produce unverifiable hallucinations, due to overlong length of speech, multi-party interactions and unstructured chatting. Hereby, we propose a StreamProfile, a streaming framework that processes counseling speech incrementally, extracts evidences grounded from ASR transcriptions by storing it in a Hierarchical Evidence Memory, and then performs a Chain-of-Thought pipeline according to PM+ psychological intervention for clinical reasoning. The final profile is synthesized strictly from those evidences, making every claim traceable. Experiments on real-world teenager counseling speech have shown that the proposed StreamProfile system can accurately generate the profiles and prevent hallucination. 
\end{abstract}


\section{Introduction}
\label{sec:intro}

Psychological counseling is a primary intervention to manage the mental health conditions, especially for adolescents, where the Problem Management Plus (PM+) developed by the WHO is commonly adopted\cite{dawson2015problem}. The core therapeutic strategies include problem solving to help individuals identify, prioritize and tackle practical problems (e.g. family conflict, learning stress, anxiety) in a structured way\cite{winters2007case, eells2022handbook, sim2005case}. Therefore, the psychological profile is essential which systematically records a patient’s background, relationships, emotional state, cognitive patterns, and risk indicators across standardized dimensions. A typical counseling session involves talking one-on-one with a trained professional, lasting 30 to 90 minutes. It is interesting to summarize automatically from the recordings, although it may include noises, far-field speech, even multi-party dialogues when the parents are involved in the teenager counseling. Moreover, the depression patients commonly have paralinguistic features, such as slowed speech, monotone voice, reduced volume, increased pauses and difficulties to express his/her feeling\cite{mundt2012vocal,alpert2001reflections}, which is a challenge to scrutinize the speech data and summarize the structured psychological profile.

Recent advances in large language models (LLMs) have demonstrated strong capabilities in long-form summarization and information extraction, making them a natural candidate for this task\cite{liu2024benchmarking, gupta2025autosumm, oliveira2025development, gong2025comparison}. A straightforward approach is to transcribe the session audio and prompt an LLM to generate the profile from the full transcript. In practice, however, this approach suffers from two compounding failure modes that are particularly consequential in a clinical setting. First, LLMs exhibit a well-known tendency to disproportionately attend to the beginning and end of long inputs, causing clinically critical details in the middle of a session to be silently dropped\cite{liu2024loss}. Second, LLMs hallucinate by generating plausible-sounding but factually unsupported claims about patient history, risk level, or interpersonal dynamics\cite{belem2025single, zhang2023extractive}. In a psychiatric workflow, a fabricated self-harm history is not merely an accuracy failure but a direct patient safety risk, and clinicians cannot act on summaries they cannot verify.

Retrieval-augmented generation (RAG) and agent memory techniques offer potential remedies by grounding generation in retrieved context\cite{hu2025memory, xu2025mem, hu2025hiagent, xu2022long}.
However, existing approaches are predominantly designed for well-structured text sources such as academic papers, product manuals, or knowledge bases, where entities and relations are explicitly stated.
Psychological counseling dialogue is fundamentally different: it is unstructured, emotionally laden, and multi-party, with clinically relevant disclosures embedded in conversational filler, indirect expression, and role-specific speech acts.
These characteristics make it difficult for standard RAG or flat memory systems to reliably identify, classify, and retain the sparse clinical signals that are essential for profile generation.

We propose \textbf{StreamProfile}, a streaming psychological profile generate framework that processes session audio in real time. The core contributions are as follows:
(1).We design a Chain-of-Thought (CoT) pipeline based on the PM+ Protocol\cite{wei2022chain} that enforces explicit clinical reasoning, performing epistemic filtering to distinguish direct patient disclosures from counselor restatements and hypothetical examples, and extracting structured evidence tuples grounded to verbatim utterances.
(2).we introduce the Hierarchical Evidence Memory (HEM), a persistent, clinically-aligned agent memory that continuously accumulates and indexes evidence from unstructured conversational speech throughout the session, ensuring that every generated claim in the final profile is traceable to a specific patient utterance.
(3).we conduct comprehensive experiments on Psy-Bench, a dataset of real-world chinese counseling sessions, demonstrating substantial improvements over LLM baselines on both profile generate performance and hallucination evaluation.

\begin{figure*}
\centering 
\includegraphics[width=1\linewidth]{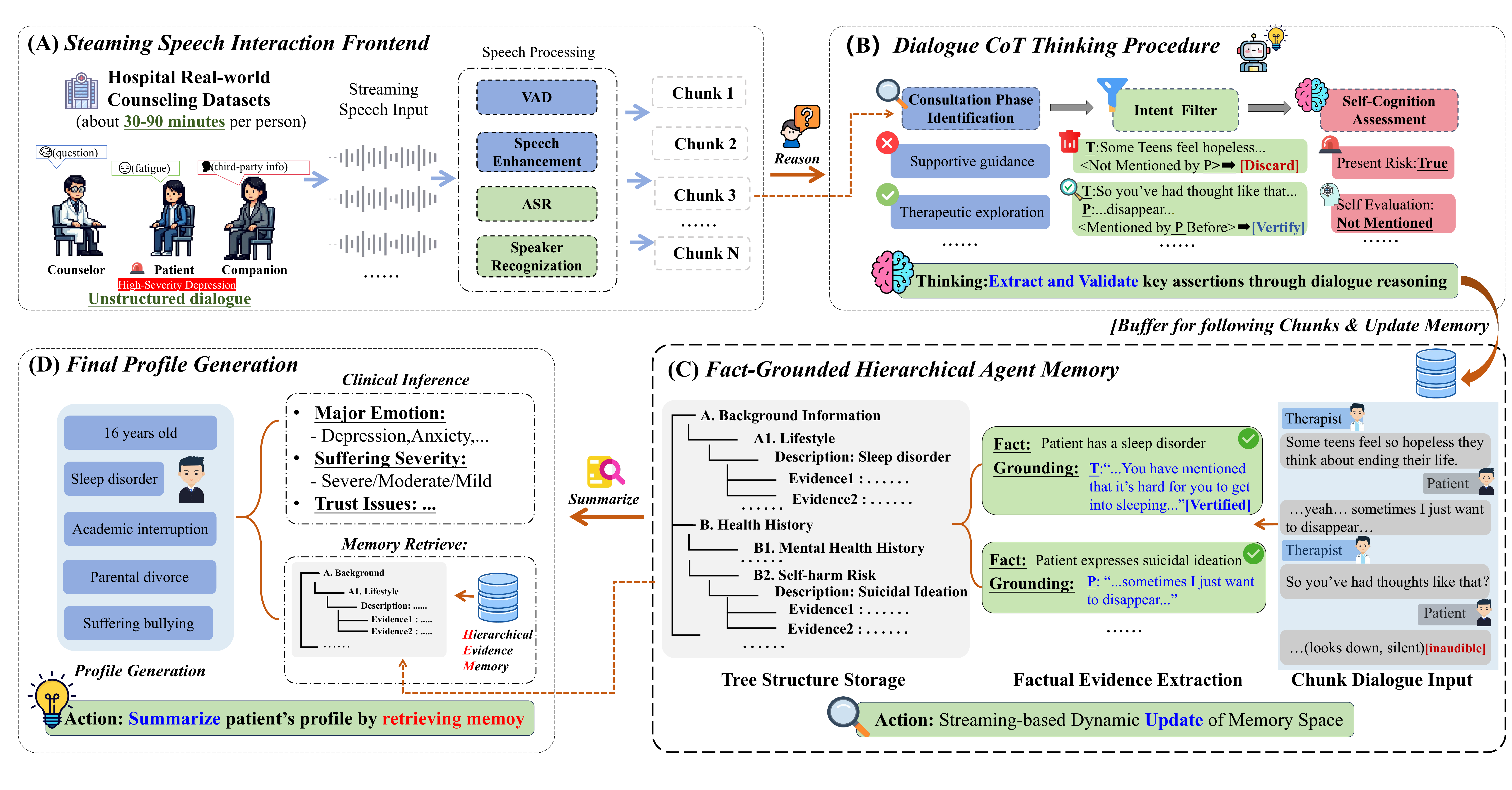} 
\caption{The StreamProfile framework processes streaming audio, performs PM+-guided CoT reasoning, and accumulates evidence in Hierarchical Evidence Memory for final profile generation.} \label{fig:framework} 
\end{figure*}


\section{Methodology}
\label{sec:method}

StreamProfile processes a raw session recording $\mathcal{A}$ in a streaming fashion, triggering a processing cycle every $T$ seconds as audio arrives.
At each cycle, the most recent $T$-second window is passed through a speech front-end to produce a diarized transcript, which is then processed by a CoT pipeline.
Extracted evidence is accumulated incrementally in a HEM, enabling continuous profile updates throughout the session.
Upon session completion, the final profile is synthesized from the fully populated HEM.The complete pipeline is summarized in Algorithm~\ref{alg:streaming_profile}.

\subsection{Speech Interaction Frontend}
\label{sec:speech}

Each $T$-second audio window $\mathcal{A}_n$ is processed by a sequential pipeline to produce a structured transcript.
The raw waveform is first enhanced by FastEnhancer\footnote{\url{https://github.com/modelscope/FunASR}}, a lightweight neural speech enhancement model that suppresses background noise and reverberation common in clinical recording environments.
Voice Activity Detection (VAD) then segments the enhanced signal into speech intervals. These intervals are transcribed by SenseVoice\cite{gao2023funasr}\footnote{\url{https://github.com/FunAudioLLM/SenseVoice}}, yielding a transcription $u_i$ for each segment $i$.
For speaker diarization, CAM++\footnote{\url{https://github.com/modelscope/3D-Speaker}} speaker embeddings are extracted per segment and clustered via spectral clustering with the speaker count constrained to $s \in \{2, 3\}$, selected by maximizing the silhouette coefficient.
Role assignment is performed via one-shot speaker verification: a short enrollment utterance from the known counselor is compared against each cluster centroid by cosine similarity, and the closest cluster is labeled as the counselor, with remaining clusters assigned to patient and guardian roles per the session protocol.
A global re-clustering pass over all accumulated embeddings at session end corrects within-window assignment drift.The output for window $n$ is a structured dialogue sequence $\mathcal{D}_n = \{(u_i,\, r_i,\, t_i^{\mathrm{s}},\, t_i^{\mathrm{e}})\}_{i=1}^{M_n}$, where $r_i \in \{\text{counselor, patient, guardian}\}$ and $[t_i^{\mathrm{s}}, t_i^{\mathrm{e}}]$ are absolute timestamps within the session.

\subsection{Chain-of-Thought Based on the PM+ Protocol}
\label{sec:cot}

The CoT pipeline emulates a clinician's stepwise reasoning based on the PM+ protocol, transforming raw dialogue into clinically meaningful evidence. For each window $\mathcal{D}_n$, the pipeline executes three reasoning steps.

\textbf{Step 1: Consultation Phase Identification.}
The model first identifies which of the three canonical consultation modules below are present in the chunk:

(1)Supportive Guidance: Includes psycho-education, relaxation techiques (e.g., diaphragmatic breathing), or general psychological facts. 

(2)Objective Evaluation: Involves structured symptom assessment, where the clinician repeatedly asks the patient to rate specific negative feelings or confirms scores on standardized scales.

(3)Problem Management: Characterized by open-ended inquiry into core life domains, such as family dynamics, peer relationships, Stress related to school or work, self-worth, and risks of self-harm. Higher analytical priority is assigned to this module, which clinically significant evidence.

\textbf{Step 2: Speaker-Intent Filtering.}
In high-severity depression, patients often speak minimally or not at all, leading to poor ASR transcription and significant loss of patient-generated content. Clinicians commonly respond with confirmatory paraphrasing, restating the patient’s fragmented or implied meaning in clearer form. Although these paraphrases accurately capture patient intent, they appear in transcripts as therapist speech, causing therapist-originated content to dominate the dialogue. To address this issue, we require the model to perform intent filtering, distinguishing therapist paraphrasing from genuine patient speech to preserve clinically valid evidence. The prompt explicitly instructs the agent to apply a speaker-intent filtering rule:
(1).Retain therapist statements that paraphrase or confirm the patient’s prior (explicit or implicit) disclosure;
(2).Discard therapist statements that present generic hypotheses, illustrative examples, or factual explanations.

\textbf{Step 3: Risk and Emotional State Inference.}
The model then performs a deep exploration of self-harm risk and self-cognition, identifying explicit or implicit signals of suicidal ideation. Crucially, it treats emotional distress as a multi-faceted construct through a two-step clinical inference:

(1)Suffering Severity Assessment: It distinguishes clinically significant distress from normative stress based on the temporal scope and functional impact of stressors and the presence of high-risk indicators.

(2)Major Emotion Categorization: It concurrently infers the dominant emotion. If the patient is clinically distressed, the label is selected from pathological categories (e.g., depression); if normative, from everyday affect (e.g., sadness).

This differentiation prevents over-pathologizing normal responses while accurately capturing clinical states. The reasoning outputs a free-text analysis $\phi_n$ summarizing the findings, which is appended to a rolling history buffer $\mathcal{H}$ for cross-window context.

\subsection{Hierarchical Evidence Memory}
\label{sec:hem}

The Hierarchical Evidence Memory (HEM) is a persistent agent memory that accumulates and structures evidence from the CoT pipeline. It ensures that every clinical claim in the final profile is grounded in a verifiable source utterance.

After CoT reasoning produces $\phi_n$, the model performs a second LLM call to extract structured evidence from the dialogue window. This extraction takes both the CoT analysis and the original transcript as input:
\[
\mathcal{E}_n = \mathrm{LLM}_{\mathrm{extract}}(\phi_n, \mathcal{D}_n)
\]
where $\mathcal{E}_n$ is a set of tuples $(d_j, a_j, s_j)$. Here $d_j$ is a clinical dimension from the predefined schema $\Omega$ (Table~\ref{tab:hem_structure}), $a_j$ is a verbatim ASR utterance copied from $\mathcal{D}_n$, and $s_j$ is a concise patient-centric insight derived from that utterance. The extraction is constrained to cover all dimensions, prioritize risk-related ones, and avoid negated statements.

Each tuple is then stored in the HEM, which organizes entries by dimension. To prevent redundancy, a new evidence $s_j$ is accepted only if its bi-gram Jaccard similarity with all existing entries in the same dimension is below a predefined threshold $T$. Each stored entry also includes the utterance’s timestamps and detected emotion, enabling full traceability. This dual storage supports hallucination-free profile synthesis: the final generation uses aggregated insights, while raw utterances remain available for clinician verification.

\begin{table}[t]
\centering
\caption{Clinical schema $\Omega$ of the psychological profile }
\label{tab:hem_structure}
\resizebox{\columnwidth}{!}{%
\begin{tabular}{llp{5.5cm}}
\toprule
\textbf{Module} & \textbf{ID} & \textbf{Sub-Dimensions} \\
\midrule
Background       & A & Lifestyle (A1), Education (A2) \\
Health History   & B & Mental Health Hx (B1), Self-harm Hx (B2) \\
Social Relations & C & Family (C1), Peers (C2), Social (C3), Academic (C4) \\
Self-Cognition   & D & Self-Perception (D1), Self-Worth (D2) \\
Current State    & E & Emotion (E1), Recent Events (E2), Depression (E3), Trust (E4), Anxiety (E5), Distress (E6), Bullying (E7) \\
\bottomrule
\end{tabular}}
\end{table}

\begin{algorithm}[t]
\caption{StreamProfile}
\label{alg:streaming_profile}
\begin{algorithmic}[1]
\REQUIRE Session audio stream $\mathcal{A}$, window size $T$, LLM $f_\theta$, schema $\Omega$
\ENSURE Profile $\mathcal{P}$, evidence memory $\mathcal{M}$
\STATE $\mathcal{M} \leftarrow \mathrm{InitHEM}(\Omega)$;\quad $\mathcal{H} \leftarrow \emptyset$
\WHILE{$true$}
  \STATE $\mathcal{D}_n \leftarrow \mathrm{FrontEnd}(\mathcal{A}_n)$ \hfill \{enhance, VAD, ASR, diarize\}
  \STATE $\phi_n \leftarrow f_\theta^{\mathrm{parse}}(\mathcal{D}_n,\, \mathcal{H})$ \hfill \{Stage 1\}
  \STATE $\mathcal{E}_n \leftarrow f_\theta^{\mathrm{extract}}(\mathcal{D}_n,\, \phi_n)$ \hfill \{Stage 2\}
  \FOR{each $(d_j, a_j, \kappa_j) \in \mathcal{E}_n$}
    \IF{$\max_{\kappa' \in \mathcal{M}(d_j)} J(\kappa_j, \kappa') \leq threshold\ T=0.7$}
      \STATE $\mathcal{M}.\mathrm{Store}(d_j, \kappa_j, a_j)$ \hfill \{Stage 3\}
    \ENDIF
  \ENDFOR
  \STATE $\mathcal{H} \leftarrow \mathcal{H} \cup \{\phi_n\}$
\ENDWHILE
\STATE $\mathcal{P} \leftarrow f_\theta^{\mathrm{profile}}(\mathcal{M}.\mathrm{Retrieve}())$
\STATE \textbf{return} $\mathcal{P}$, $\mathcal{M}$
\end{algorithmic}
\end{algorithm}

\section{Experimental}

\subsection{Experimental Setup}
\label{sec:setup}
We used different LLMs as comparison methods, with all generation parameters configured identically to ensure fairness: temperature was set to 0.3, maximum output length to 8,192 tokens, and random seed to 42. For API-based models, inference was conducted via their official APIs with the same generation parameters. For local LLMs, we used the vLLM framework with tensor parallelism across 8 $\times$ NVIDIA RTX 5880 GPUs, GPU memory utilization set to 0.9, and a maximum context length of 32,768 tokens.


\subsection{Dataset}
We evaluate on \textbf{Psy-Bench}, a real-world Chinese psychological counseling dataset collected in hospital psychiatric outpatient and community settings under the PM+ protocol.It contains 19 annotated sessions from three participant groups: (i) clinically diagnosed adolescent depression patients, (ii) suspected individuals presenting anxiety or sub-threshold symptoms.Sessions range from 30 to 90 minutes and involve a counselor, a patient, and in some cases a guardian.
All ground-truth profiles are independently annotated by three PhD-level psychologists following the clinical schema $\Omega$ defined in Table~\ref{tab:hem_structure}, with disagreements resolved by consensus.

\subsection{Evaluation metrics.}
All generated profiles are evaluated against reference profiles manually constructed by psychologists. These expert annotations serve as the clinical ground truth, ensuring our metrics reflect clinically meaningful. For objective NLP quality, We compute ROUGE-L\cite{lin2004rouge}, BERTScore\cite{zhang2019bertscore}, SBERT\cite{reimers2019sentence} cosine similarity score, where the ground-truth annotation serves as the premise and the generated claim as the hypothesis. For hallucination evaluation, a LLM-based judge scores 3 clinical quality dimensions on a 1--5 Likert scale: Coverage measures how many reference key facts are captured in the generated profile; Consistency measures whether every generated claim is grounded in the session transcript; and Hallucination penalizes fabricated or unverifiable content, including claims that appear plausible but are absent from the source dialogue.

\begin{figure}[t]
\centering
\includegraphics[width=\linewidth]{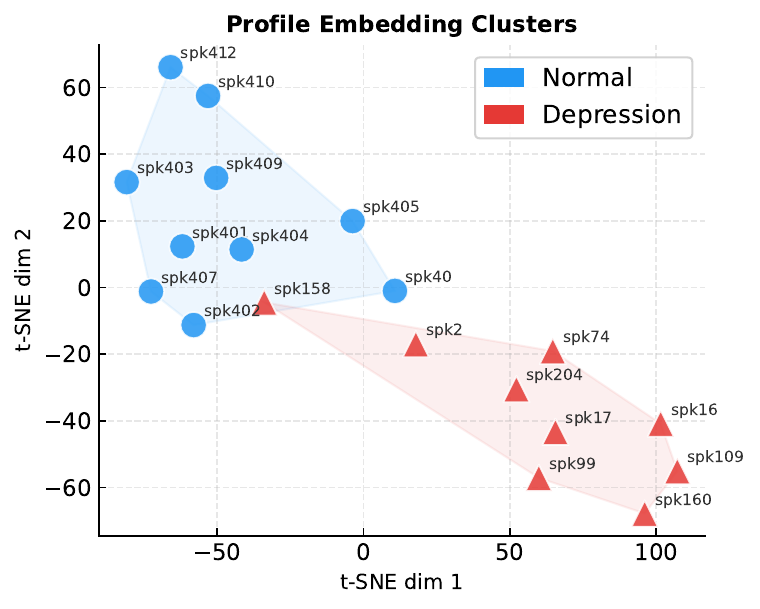}
\caption{t-SNE projection of profile embeddings generated by StreamProfile (DeepSeek-CoT). }
\label{fig:cluster}
\end{figure}

\begin{figure}[t]
\centering
\includegraphics[width=\linewidth]{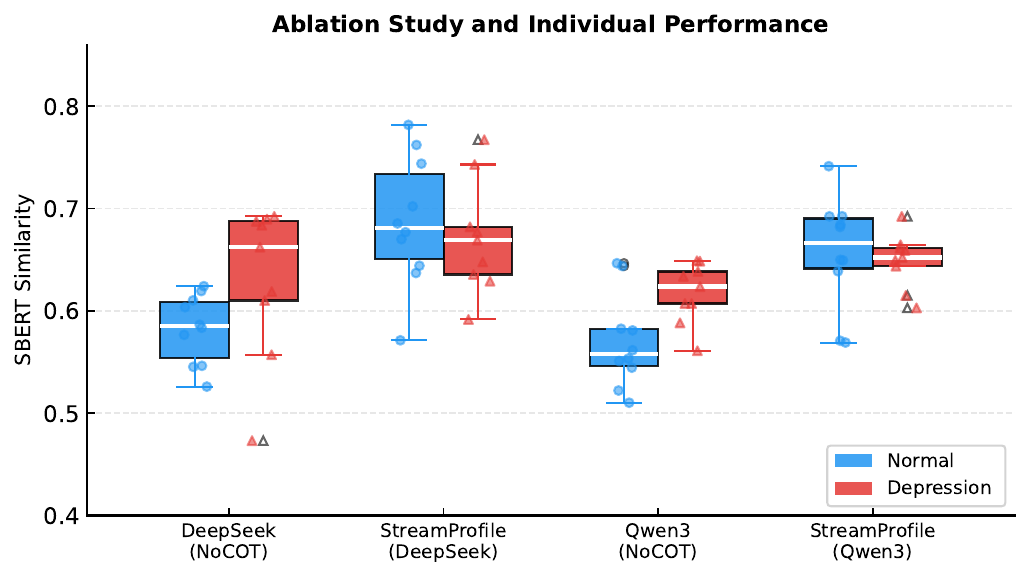}
\caption{Ablation Study: Based on SBERT-Scores distribution across sessions for four conditions, grouped by patient type (Normal vs.~depression).}
\label{fig:sbert_session}
\end{figure}

\subsection{Results on Profile Generation}
\label{sec:results_nlp}

Table~\ref{tab:profile_comparison} reports NLP similarity results.
StreamProfile consistently achieves the best scores across all metrics under both backbone LLMs. StreamProfile (DeepSeek) reaches SBERT 0.680, ROUGE-L 0.216, and BERTScore-F 0.702, surpassing the strongest text-LLM baseline in average. Furthermore, StreamProfile (Qwen3) likewise outperforms its No-CoT counterpart by 0.051, confirming that the CoT and HEM framework contributes regardless of the backbone.

Among baselines, Qwen2-Audio scores lowest because it must simultaneously perform speech recognition and clinical reasoning in a single forward pass, leading to content truncation and sparse output. For LLaMA-3.1-70B, Despite its 70B parameter scale, its extracts only surface-level facts, omitting clinically critical details like self-harm history, indicating that model scale can producted limited multilingual clinical reasoning.

For ablation Study, Figure~\ref{fig:sbert_session} further illustrates that CoT variants consistently outperform No-COT conditions across both patient groups, with the advantage more pronounced for depression sessions where clinically relevant signals are sparse and require multi-step reasoning to surface. Furthermore, Figure~\ref{fig:cluster} shows that profiles generated by StreamProfile form well-separated clusters for depression patients and non-clinical Normal in t-SNE space, even though no group label is provided at generation time.

\begin{table}[t]
\centering
\caption{Profile similarity comparison between different models and our proposed system against human-annotated reference profiles, where $^\dagger$ denotes the baseline}
\label{tab:profile_comparison}
\resizebox{\columnwidth}{!}{
\begin{tabular}{l|cccc}
\toprule
\textbf{Model} & \textbf{SBERT$\uparrow$} & \textbf{ROUGE-L$\uparrow$} & \textbf{BERTScore-F$\uparrow$} & \textbf{Average$\uparrow$} \\
\midrule
\multicolumn{5}{c}{\textit{Audio LLM}} \\
\midrule
Qwen2-Audio~\cite{chu2024qwen2} & 0.323 & 0.056 & 0.481 & 0.287 \\
\midrule
\midrule
\multicolumn{5}{c}{\textit{Text LLM (No COT)}} \\
\midrule
Qwen3-252B\cite{yang2025qwen3}$^\dagger$        & 0.592 & 0.132 & 0.648 & 0.457 \\
LLaMA-3.1-70B\cite{grattafiori2024llama}     & 0.477 & 0.089 & 0.605 & 0.391 \\
DeepSeek-V3~\cite{guo2025deepseek}$^\dagger$ & 0.605 & 0.136 & 0.648 & 0.463 \\
\midrule
\midrule
\multicolumn{5}{c}{\textit{Ours (StreamProfile)}} \\
\midrule
\textbf{StreamProfile (Qwen3)}    & \textbf{0.653} & \textbf{0.187} & \textbf{0.685} & \textbf{0.508} \\
\textbf{StreamProfile (DeepSeek)} & \textbf{0.680} & \textbf{0.216} & \textbf{0.702} & \textbf{0.533} \\
\bottomrule
\end{tabular}
}
\end{table}

\subsection{Results on Hallucination and Clinical Fidelity}
\label{sec:results_llm}

Table~\ref{tab:hall_subjective} reports LLM-as-Judge scores.
StreamProfile (DeepSeek) achieves the highest scores on all three dimensions, improving over DeepSeek-V3(No-CoT) despite using the same backbone, confirming that gains come from the CoT and HEM framework. The most notable improvement appears on hallucination: a gain of 0.39 over No-CoT, because profile synthesis is strictly conditioned on HEM entries each grounded to a verbatim utterance, preventing the model from generating unsupported clinical claims. Furthermore, the streaming CoT design further enhances consistency by preserving reasoning continuity, mitigating the lost-in-the-middle problem that causes No-CoT baselines to produce fragmented or self-contradictory outputs.

Neverthless, LLaMA-3.1-70B scores 2.46 on hallucination yet only 1.21 on coverage, revealing that it avoids fabrication by producing sparse, non-committal profiles rather than accurate ones, a failure mode that is equally unusable in clinical practice.Qwen2-Audio scores lowest overall, with coverage of only 1.15, as its end-to-end audio pipeline conflates acoustic noise with clinical content and generates outputs faithful to neither the reference nor the transcript.

\begin{table}[t]
\centering
\caption{LLM-as-judge evaluation of three dimensions (Hallucination, Coverage, Consistency) on a 1–5 scale.}
\label{tab:hall_subjective}
\resizebox{\columnwidth}{!}{
\begin{tabular}{l|l|cccc}
\toprule
\multirow{2}{*}{\textbf{Type}} & \multirow{2}{*}{\textbf{Model}} & \multicolumn{4}{c}{\textbf{LLM-As-Judge}} \\
\cmidrule(lr){3-6}
& & Hallucination & Coverage & Consistency & Average \\
\midrule
\multirow{1}{*}{Audio} 
& Qwen2-Audio\cite{chu2024qwen2} & 2.02 & 1.15 & 1.56 & 1.58 \\
\midrule
\multirow{3}{*}{Text} 
& LLaMA-3.1-70B\cite{grattafiori2024llama}    & 2.46 & 1.21 & 1.89 & 1.86 \\
& Qwen3-252B\cite{yang2025qwen3}       & 2.38 & 2.38 & 2.71 & 2.49 \\
& DeepSeek-V3\cite{guo2025deepseek}      & 2.43 & 2.52 & 2.84 & 2.60 \\
\midrule
\multirow{2}{*}{Ours}  
& \textbf{StreamProfile (Qwen3)}    & \textbf{2.54} & \textbf{2.53} & \textbf{2.86} & \textbf{2.64} \\
& \textbf{StreamProfile (DeepSeek)} & \textbf{2.82} & \textbf{2.63} & \textbf{3.05} & \textbf{2.83} \\
\bottomrule
\end{tabular}
}
\end{table}

\section{Conclusion}

We introduced StreamProfile, a streaming framework for generating structured psychological profiles from counseling sessions. By integrating a CoT reasoning pipeline with a HEM, our approach ensures that every clinical claim is grounded in verifiable utterances, effectively mitigating long-context forgetting and hallucination. Experiments on real-world data demonstrate that StreamProfile significantly outperforms LLM baselines in both profile accuracy and clinical fidelity, providing a transparent pathway for trustworthy AI assistance in mental healthcare.





\section{Generative AI Use Disclosure}
Generative AI tools were used solely for manuscript language polishing. They were not used to create any core research content, results, or arguments. All authors are fully responsible for the work and consent to its submission. No generative AI tool is a co-author.

\bibliographystyle{IEEEtran}
\bibliography{mybib}

@article{hu2025memory,
  title={Memory in the age of ai agents},
  author={Hu, Yuyang and Liu, Shichun and Yue, Yanwei and Zhang, Guibin and Liu, Boyang and Zhu, Fangyi and Lin, Jiahang and Guo, Honglin and Dou, Shihan and Xi, Zhiheng and others},
  journal={arXiv preprint arXiv:2512.13564},
  year={2025}
}

@article{xu2025mem,
  title={A-mem: Agentic memory for llm agents},
  author={Xu, Wujiang and Liang, Zujie and Mei, Kai and Gao, Hang and Tan, Juntao and Zhang, Yongfeng},
  journal={arXiv preprint arXiv:2502.12110},
  year={2025}
}

@inproceedings{hu2025hiagent,
  title={Hiagent: Hierarchical working memory management for solving long-horizon agent tasks with large language model},
  author={Hu, Mengkang and Chen, Tianxing and Chen, Qiguang and Mu, Yao and Shao, Wenqi and Luo, Ping},
  booktitle={Proceedings of the 63rd Annual Meeting of the Association for Computational Linguistics (Volume 1: Long Papers)},
  pages={32779--32798},
  year={2025}
}

@inproceedings{zhang2023extractive,
  title={Extractive summarization via chatgpt for faithful summary generation},
  author={Zhang, Haopeng and Liu, Xiao and Zhang, Jiawei},
  booktitle={Findings of the association for computational linguistics: EMNLP 2023},
  pages={3270--3278},
  year={2023}
}

@inproceedings{belem2025single,
  title={From single to multi: How llms hallucinate in multi-document summarization},
  author={Belem, Catarina G and Pezeshkpour, Pouya and Iso, Hayate and Maekawa, Seiji and Bhutani, Nikita and Hruschka, Estevam},
  booktitle={Findings of the Association for Computational Linguistics: NAACL 2025},
  pages={5276--5309},
  year={2025}
}

@inproceedings{liu2024benchmarking,
  title={Benchmarking generation and evaluation capabilities of large language models for instruction controllable summarization},
  author={Liu, Yixin and Fabbri, Alexander Richard and Chen, Jiawen and Zhao, Yilun and Han, Simeng and Joty, Shafiq and Liu, Pengfei and Radev, Dragomir and Wu, Chien-Sheng and Cohan, Arman},
  booktitle={Findings of the Association for Computational Linguistics: NAACL 2024},
  pages={4481--4501},
  year={2024}
}

@inproceedings{liu2024loss,
  title={Lost in the Middle: How Language Models Use Long Contexts},
  author={Liu, Nelson F. and Lin, Kevin and Hewitt, John and Paranjape, Ashwin and Bevilacqua, Michele and Petroni, Fabio and Liang, Percy},
  booktitle={Transactions of the Association for Computational Linguistics},
  volume={12},
  pages={157--173},
  year={2024},
  publisher={MIT Press},
  doi={10.1162/tacl_a_00638},
  url={https://aclanthology.org/2024.tacl-10/}
}

@article{guo2025deepseek,
  title={DeepSeek-R1 incentivizes reasoning in LLMs through reinforcement learning},
  author={Guo, Daya and Yang, Dejian and Zhang, Haowei and Song, Junxiao and Wang, Peiyi and Zhu, Qihao and Xu, Runxin and Zhang, Ruoyu and Ma, Shirong and Bi, Xiao and others},
  journal={Nature},
  volume={645},
  number={8081},
  pages={633--638},
  year={2025},
  publisher={Nature Publishing Group UK London}
}

@inproceedings{lin2004rouge,
  title={Rouge: A package for automatic evaluation of summaries},
  author={Lin, Chin-Yew},
  booktitle={Text summarization branches out},
  pages={74--81},
  year={2004}
}

@article{zhang2019bertscore,
  title={Bertscore: Evaluating text generation with bert},
  author={Zhang, Tianyi and Kishore, Varsha and Wu, Felix and Weinberger, Kilian Q and Artzi, Yoav},
  journal={arXiv preprint arXiv:1904.09675},
  year={2019}
}

@inproceedings{xu2022long,
    title = "Long Time No See! Open-Domain Conversation with Long-Term Persona Memory",
    author = "Liu, Yibin  and Xu, Weidi  and Zhou, Hao  and Wan, Xiaojun  and Huang, Minlie",
    editor = "Muresan, Smaranda  and Yap, Georgiana  and Fung, Pascale",
    booktitle = "Findings of the Association for Computational Linguistics: ACL 2022",
    month = may,
    year = "2022",
    address = "Dublin, Ireland",
    publisher = "Association for Computational Linguistics",
    url = "https://aclanthology.org/2022.findings-acl.207",
    doi = "10.18653/v1/2022.findings-acl.207",
    pages = "2663--2676"
}

@inproceedings{reimers2019sentence,
  title={Sentence-bert: Sentence embeddings using siamese bert-networks},
  author={Reimers, Nils and Gurevych, Iryna},
  booktitle={Proceedings of the 2019 conference on empirical methods in natural language processing and the 9th international joint conference on natural language processing (EMNLP-IJCNLP)},
  pages={3982--3992},
  year={2019}
}

@article{chu2024qwen2,
  title={Qwen2-audio technical report},
  author={Chu, Yunfei and Xu, Jin and Yang, Qian and Wei, Haojie and Wei, Xipin and Guo, Zhifang and Leng, Yichong and Lv, Yuanjun and He, Jinzheng and Lin, Junyang and others},
  journal={arXiv preprint arXiv:2407.10759},
  year={2024}
}

@article{alpert2001reflections,
  title={Reflections of depression in acoustic measures of the patient’s speech},
  author={Alpert, Murray and Pouget, Enrique R and Silva, Raul R},
  journal={Journal of affective disorders},
  volume={66},
  number={1},
  pages={59--69},
  year={2001},
  publisher={Elsevier}
}

@article{grattafiori2024llama,
  title={The llama 3 herd of models},
  author={Grattafiori, Aaron and Dubey, Abhimanyu and Jauhri, Abhinav and Pandey, Abhinav and Kadian, Abhishek and Al-Dahle, Ahmad and Letman, Aiesha and Mathur, Akhil and Schelten, Alan and Vaughan, Alex and others},
  journal={arXiv preprint arXiv:2407.21783},
  year={2024}
}

@article{mundt2012vocal,
  title={Vocal acoustic biomarkers of depression severity and treatment response},
  author={Mundt, James C and Vogel, Adam P and Feltner, Douglas E and Lenderking, William R},
  journal={Biological psychiatry},
  volume={72},
  number={7},
  pages={580--587},
  year={2012},
  publisher={Elsevier}
}

@article{gao2023funasr,
  title={Funasr: A fundamental end-to-end speech recognition toolkit},
  author={Gao, Zhifu and Li, Zerui and Wang, Jiaming and Luo, Haoneng and Shi, Xian and Chen, Mengzhe and Li, Yabin and Zuo, Lingyun and Du, Zhihao and Xiao, Zhangyu and others},
  journal={arXiv preprint arXiv:2305.11013},
  year={2023}
}

@article{yang2025qwen3,
  title={Qwen3 technical report},
  author={Yang, An and Li, Anfeng and Yang, Baosong and Zhang, Beichen and Hui, Binyuan and Zheng, Bo and Yu, Bowen and Gao, Chang and Huang, Chengen and Lv, Chenxu and others},
  journal={arXiv preprint arXiv:2505.09388},
  year={2025}
}

@inproceedings{gong2025comparison,
  title={Comparison-Based Automatic Evaluation for Meeting Summarization},
  author={Gong, Ziwei and Ai, Lin and Deshpande, Harsh and Johnson, Alexander and Phung, Emmy and Wu, Zehui and Emami, Ahmad and Hirschberg, Julia},
  booktitle={Proc. Interspeech 2025},
  pages={291--295},
  year={2025}
}

@article{sim2005case,
  title={Case formulation in psychotherapy: Revitalizing its usefulness as a clinical tool},
  author={Sim, K. and Gwee, K. P. and Bateman, A.},
  journal={Academic Psychiatry},
  volume={29},
  number={3},
  pages={289--292},
  year={2005}
}

@book{eells2022handbook,
  title={Handbook of psychotherapy case formulation},
  author={Eells, Tracy D},
  year={2022},
  publisher={Guilford Publications}
}

@article{winters2007case,
  title={The case formulation in child and adolescent psychiatry},
  author={Winters, Nancy C and Hanson, Graeme and Stoyanova, Veneta},
  journal={Child and Adolescent Psychiatric Clinics},
  volume={16},
  number={1},
  pages={111--132},
  year={2007},
  publisher={Elsevier}
}

@article{oliveira2025development,
  title={Development and evaluation of a clinical note summarization system using large language models},
  author={Oliveira, J. D. and Santos, H. D. P. and Ulbrich, A. H. D. P. S. and others},
  journal={Communications Medicine},
  volume={5},
  number={1},
  pages={376},
  year={2025}
}

@inproceedings{gupta2025autosumm,
  title={AUTOSUMM: A comprehensive framework for LLM-based conversation summarization},
  author={Gupta, Abhinav and Singh, Devendra and Cowan, Greig A and Kadhiresan, N and Srivastava, Siddharth and Sriraja, Yagneswaran and Mantri, Yoages Kumar},
  booktitle={Proceedings of the 63rd Annual Meeting of the Association for Computational Linguistics (Volume 6: Industry Track)},
  pages={500--509},
  year={2025}
}

@article{wei2022chain,
  title={Chain-of-thought prompting elicits reasoning in large language models},
  author={Wei, Jason and Wang, Xuezhi and Schuurmans, Dale and Bosma, Maarten and Xia, Fei and Chi, Ed and Le, Quoc V and Zhou, Denny and others},
  journal={Advances in neural information processing systems},
  volume={35},
  pages={24824--24837},
  year={2022}
}

@article{dawson2015problem,
  title={Problem Management Plus (PM+): a WHO transdiagnostic psychological intervention for common mental health problems},
  author={Dawson, Katie S and Bryant, Richard A and Harper, Melissa and Tay, Alvin Kuowei and Rahman, Atif and Schafer, Alison and Van Ommeren, Mark},
  journal={World Psychiatry},
  volume={14},
  number={3},
  pages={354},
  year={2015}
}

\end{document}